\documentclass{an}
\def\gtsim{\hbox{\raise 2pt \hbox {$>$} \kern-1.1em \lower 4pt \hbox {$\sim$}}}
\usepackage{graphicx}
\usepackage{times}
\usepackage{fancyhdr}
\begin{document}

\title{Observations of magnetic fields in regular and irregular clusters}

\author{Federica Govoni\inst{1}}
\institute{
INAF - Osservatorio Astronomico di Cagliari, Loc. Poggio dei Pini,
Strada 54, 09012 Capoterra (CA), Italy\\
INAF - Istituto di Radioastronomia, Via Gobetti 101,
I-40129 Bologna, Italy
}

\date{Received; accepted; published online}

\abstract{The existence of magnetic fields associated with the intracluster
medium in clusters of galaxies is now well established through
different methods of analysis.
The most detailed evidence for the presence of cluster magnetic fields 
comes from radio observations. 
Magnetic fields can be investigated through the synchrotron 
emission of cluster-wide diffuse sources 
and from studies of the rotation measure of polarized radio
galaxies.
I will review efforts to measure magnetic field strengths and power 
spectra and
the main issues that have led to our knowledge on magnetic fields in
regular and irregular clusters of galaxies.
}

\correspondence{fgovoni@ira.inaf.it}

\maketitle

\section{Introduction}
In the hierarchical scenario of structure formation, 
clusters of galaxies are formed by gravitational mergers of smaller
units as groups and sub-clusters.
The presence of substructures in the X-ray images is
indication of cluster mergers and is typical for irregular 
clusters.
Regular clusters (or cooling core clusters) are instead dynamically 
relaxed systems that don't show the presence of recent mergers.
On large scale they have a relatively smooth and symmetric 
X-ray morphology. At the center they show strong central 
X-ray peaks associated with high density and cooler gas.

Since the discovery that the intracluster medium is magnetized, 
several efforts have been directed to determine the effective 
strength and structure of cluster magnetic fields 
(see e.g. recent reviews by Carilli \& Taylor 2002; Govoni \& Feretti 2004;  
and references therein).

The most detailed evidence for the presence of cluster magnetic fields 
comes from radio observations. 
Magnetic fields are investigated through the synchrotron 
emission of cluster-wide diffuse sources (radio halos, relics and mini-halos)
and from studies of the rotation measure of polarized radio
galaxies.
Other techniques, not discussed in this review, include 
inverse Compton hard X-ray emission, cold fronts 
and magneto-hydrodynamic simulations.

I will show that it is possible to 
constrain the magnetic field strength and structure (magnetic field
power spectrum), both in regular and irregular clusters,
and I will discuss the relation between the 
magnetic field strength and the intracluster gas density.

Throughout this review I adopt the $\Lambda$CDM cosmology with
$H_0$=~71~km~s$^{-1}$Mpc$^{-1}$,
$\Omega_m$=~0.3, and $\Omega_{\Lambda}$=~0.7.

\section{Halos, relics and mini-halos}

The presence of magnetic fields in clusters is directly demonstrated
by the existence of large-scale diffuse synchrotron sources, that have
no apparent connection to any individual cluster galaxy and are
therefore associated with the intracluster medium.  
These radio sources have been
classified as radio halos, relics and mini-halos depending on their
morphology and location.

Radio halos, as revealed in Coma and some other
clusters (e.g. Giovannini \& Feretti 2002), are low surface brightness
 sources
($\simeq$10$^{-6}$ Jy arcsec$^{-2}$ at 1.4 GHz) 
located at the cluster center with a typical size of 1 Mpc and a steep 
radio spectrum\footnote{$S(\nu)\propto \nu^{-\alpha}$}
($\alpha$ \gtsim 1).

Relic sources are similar to halos but located far from the clusters center.
A spectacular example of two almost symmetric relics in the same cluster
is found in A3667 (R\"ottgering et al. 1997; Johnston-Hollitt 2003).
 
\begin{figure}
\resizebox{\hsize}{!}
{\includegraphics[]{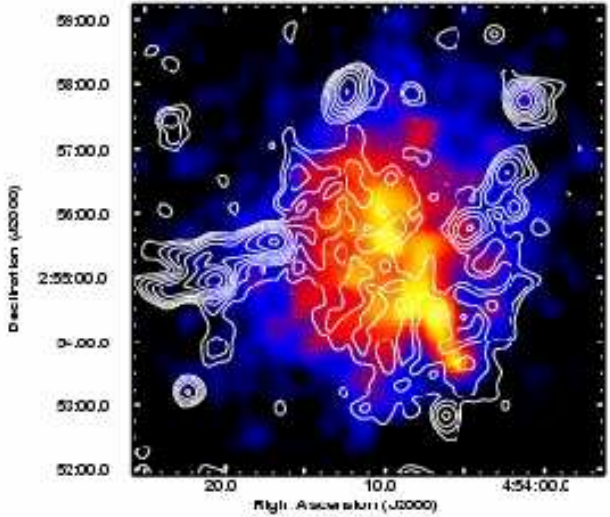}}
\caption{
The isocontour map at 1.4 GHz of the cluster A520 overlay on the
Chandra X-ray image (Markevitch et al. 2005). 
}
\label{a520}
\end{figure}

Radio halos and relics are usually present in irregular clusters
and their presence seems strictly related to the cluster merger processes, 
which can provide the energy for  the reacceleration of the relativistic electrons and the amplification of magnetic fields on
large scales. Fig.~\ref{a520} shows a comparison of the radio halo
emission (contours) in A520 overlaid on the Chandra X-ray image (gray-scale)
of the cluster (Markevitch et al. 2005).

In a few relaxed clusters
(e.g. Perseus: Burns et al. 1992; Virgo:
Owen, Eilek \& Kassim 2000), 
relativistic electrons and magnetic 
fields can be traced out quite far from the central galaxy, forming what is 
called a mini-halo.
Mini-halos are smaller than radio halos and relics. They typically
extend up to about 500 kpc from the dominant radio galaxy 
at the cluster center.

Under the assumption that radio sources are in a minimum energy condition, 
it is possible to derive a zero--order estimate of the magnetic field 
strength averaged over the entire source volume.
In the classical equipartition assumption, 
considering a power-law spectrum for the electrons, 
the magnetic field can be determined from the radio synchrotron 
luminosity and the source volume. 
These calculations typically assume a magnetic field entirely
filling the radio source, equal energy in relativistic 
protons and electrons, and a range of frequencies in which the
synchrotron luminosity is calculated from 10 MHz to 10 GHz.  
Typical equipartition magnetic fields, estimated 
in clusters with wide diffuse radio sources, are 0.1-2 $\mu$G 
(e.g. En{\ss}lin et al. 1998; Govoni et al. 2001a; Bacchi et al. 2003).
These magnetic field strengths are consistent with those 
suggested from the few detections of inverse Compton hard 
X-ray emission in clusters
containing a radio halo (e.g.
Rephaeli, Gruber \& Blanco 1999; Fusco-Femiano et al. 1999;
Rephaeli \& Gruber 2002; Fusco-Femiano et al. 2004).
However, we note that with a magnetic field
of 1 $\mu$G, for example, the electrons with  
Lorentz factor $\gamma\sim1.5\times10^3$ emit at 10 MHz, 
and thus in the equipartition approximation the contribution
of electrons with lower $\gamma$ 
is not taken into account in the total energy density.
A more accurate approach is given by adopting equipartition equations 
with a low-energy cut--off in the particle energy distribution
rather than a low-frequency cut--off in the emitted 
synchrotron spectrum 
(e.g. Brunetti, Setti \& Comastri 1997; Beck \& Krause 2005).

\section{Rotation measure}

 A complementary set of information on cluster magnetic fields can be 
derived by studying the Faraday rotation of radio galaxies in clusters.
The polarized synchrotron radiation incoming from radio sources, 
undergoes the following rotation of the
plane of polarization as it passes through the magnetized and ionized 
intracluster medium:
\begin{equation}
 \Psi_{Obs}(\lambda)=\Psi_{Int}+\lambda ^2 \times RM
\label{psi}
\end{equation}
\noindent
where $\Psi_{Obs}(\lambda)$ is the observed polarization angle
at a wavelength $\lambda$ and  $\Psi_{Int}$ is the 
intrinsic polarization angle. 

The rotation measure RM is related to the thermal electron 
density, $n_{\rm e}$, 
and magnetic field along the line-of-sight, $B_{\|}$, 
through the cluster by:
\begin{equation}
 RM = 812\int\limits_0^L n_{\rm e} B_{\|} d{\rm l} ~~~ {\rm rad~m^{-2}}
\label{rm}
\end{equation}
\noindent
where $B_{\|}$ is measured in $\mu$G, $n_{\rm e}$
in cm$^{-3}$, and $L$ is the depth of the screen in kpc.   

 Polarized radio galaxies
can be mapped at several 
frequencies to produce, by fitting Eq.~\ref{psi}, detailed RM images.
If we know the distribution of the thermal electrons in the intracluster 
medium and if the fluctuations in $n_{\rm e}$ and in $B$ are uncorrelated, 
Eq.~\ref{rm} can be used to estimate the magnetic field strength along
 the line of sight in the location of the radio galaxies.
In case of correlation between the fluctuations there may be a bias
(Beck et al. 2003).

It is worth mentioning that the RM observed in radio galaxies could be not
completely representative of the cluster magnetic field
 if the RM gets locally enhanced by the intracluster medium compression due to the motion of the radio galaxy itself.
There are, however, several RM statistical works
against this interpretation (e.g. Lawler \& Dennison 1982;
Vall\'ee, MacLeod \& Broten 1986; Kim, Tribble \& Kronberg 1991; 
Clarke, Kronberg \& B\"ohringer 2001;
Johnston-Hollit \& Ekers 2004).
In particular the statistical RM investigation 
of point sources performed by Clarke et al. (2001) show a clear
broadening of the RM distribution toward small
impact parameters indicating that most of the RM contribution comes
from the intracluster medium. Even more important,
the RM distribution of the cluster embedded sources,
which may have the problem of the local enhancement, show  
similar values of the background sources (Clarke 2004).

\begin{figure}
\resizebox{\hsize}{!}
{\includegraphics[]{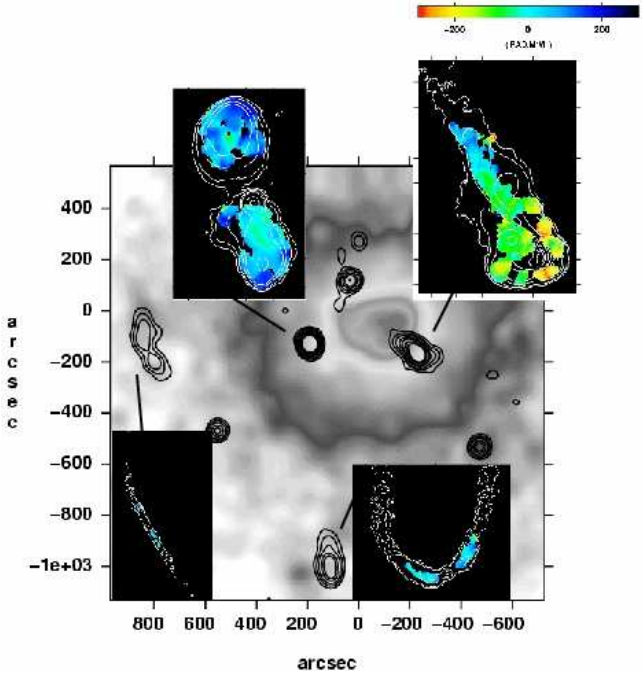}}
\caption{
NVSS contour plot and RM images (insets) 
of four radio galaxies in cluster A2255. 
The contour plot is overlaid 
onto the ROSAT X-ray image (gray-scale) of the cluster.
The RM images show fluctuations on small scales.
The RM images have been obtained by using the PACERMAN algorithm by 
Dolag, Vogt \& En{\ss}lin (2005).
}
\label{a2255}
\end{figure}

Several high quality RM images of extended radio galaxies 
exist both for regular and irregular clusters.
One example is the RM map of the radio source at the center
of the regular cluster Hydra~A (Taylor \& Perley 1993).
The image shows high RM values (thousands rad/m$^2$),
usually present at the center of cooling core clusters, where the
gas density is high.
Another beautiful example is 3C75 at the center of the irregular cluster A400
(Eilek \& Owen 2002) where the RM values are high (hundreds rad/m$^2$) 
but not so high as at the center of cooling cores clusters.
The ideal case to study the cluster magnetic field is to
sample the RM in a number as large as possible
of extended radio galaxies located in the same cluster of galaxies
(Feretti et al. 1999; Govoni et al. 2001b; Taylor et al. 2001).
Fig.~\ref{a2255} shows RM images obtained for four 
radio galaxies in the irregular cluster A2255 
(Govoni et al. in preparation). 
A common feature (visible also in Fig.~\ref{a2255}) 
of both regular and irregular clusters,
is the presence of significant variation of RM showing patchy
structures of 1-10 kpc in size.
These patchy structures indicate
that the magnetic field is tangled on small scales.

The effect of Faraday rotation from a tangled magnetic field has been
analyzed by several authors (e.g. Lawler \& Dennison 1982; Tribble
1991; Feretti et al. 1995; Felten 1996; Sokoloff et al. 1998), 
in the simplest approximation
that the magnetic field is tangled on a single scale $\Lambda_{\rm c}$.
In this ideal case, the screen is made of cells of uniform size, 
electron density
and magnetic field strength, but with a field orientation at random
angles in each cell. The observed RM along any given line of sight is
then generated by a random walk process involving a large
number of cells of size $\Lambda_{\rm c}$.  The distribution of the RM
is Gaussian with zero $<$RM$>$, and a variance given by:
\begin{equation}
 \sigma_{\rm RM}^{2}= \langle {\rm RM^{2}} \rangle = 812^{2} \Lambda_{\rm c} \int ( n_{\rm e} B_{\|})^{2}
{\rm d}l~
\label{sigmarm}
\end{equation}
\noindent
The observing strategy to get information on the cluster magnetic
field intensity and structure is to obtain high resolution 
RM maps of extended radio galaxies located in a cluster.
In case $\Lambda_{\rm c}$ is known,
the RM values combined with measurements of the thermal 
gas density can be used to derive
information on the magnetic field along line of sights
crossing different regions of the cluster.
In this approximation the magnetic field strength at the center
of irregular clusters is about 5 $\mu$G while in 
regular clusters is about 10-30 $\mu$G.

It is evident from Eq.~\ref{sigmarm} that the magnetic
field depends critically on the density of the intracluster medium and
less on the scale of the magnetic field fluctuations.
I will discuss in more detail the dependence of the magnetic field 
on these two parameters in the next subsections.

\subsection{Magnetic field and gas density connection}

By using a set of radio sources embedded 
in few irregular clusters, Dolag et al. (2001) reported a correlation between 
the cluster X-ray surface brightness and the Faraday rotation 
measure calculated in the sources location.
The analyzed clusters have similar temperatures ($\approx$ 5keV) and
follow the same correlation, as predicted by cosmological,
magneto hydrodynamical simulations (Dolag, Bartelmann \& Lesch 1999). 

In particular they found that the
$\sigma_{\rm RM}$, connected to $B$ and $n_{\rm e}$ 
(see Eq.~\ref{sigmarm}),
increases linearly with the X-ray surface brightness:
\begin{equation}
S_{\rm X} \propto \int n_{\rm e}^2 \sqrt{T} {\rm d}l
\label{xray}
\end{equation}
\noindent
The two observables $\sigma_{\rm RM}$ and S$_{\rm X}$ 
relate the two line of sight integrals 
(Eq.~\ref{sigmarm} and \ref{xray})
with each other, therefore in comparing these two quantities, 
we actually compare
cluster magnetic field versus thermal density (assuming 
a constant temperature for the whole cluster).
Thus the magnetic field profile can be represented by:
\begin{equation}
B(r)\propto n_{\rm e}(r)^\eta
\label{kingB}
\end{equation}
\noindent 
For the cluster A119, where the polarization properties of
three extended radio galaxies are available (Feretti et al. 1999), 
the $\sigma_{\rm RM}$ -- S$_{\rm X}$ relation  yields $B \propto
n_{\rm e}^{\eta}$ with $\eta=0.9$.

\begin{figure}
\resizebox{\hsize}{!}
{\includegraphics[]{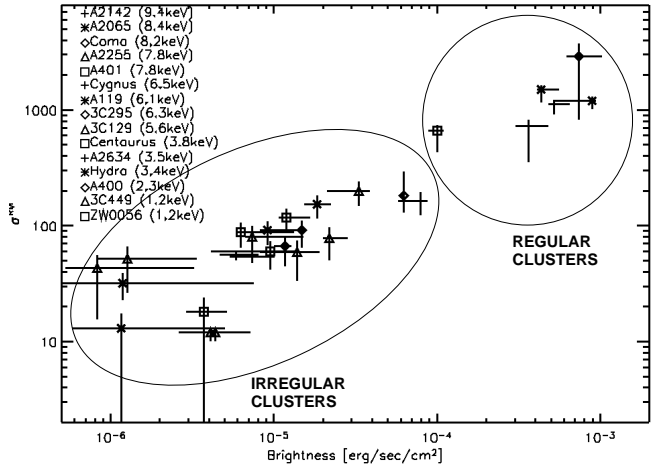}}
\caption{
Correlation between the observed $\sigma_{RM}$ 
and the X-Ray surface brightness, both for regular and irregular clusters 
(Dolag et al., in preparation).
}
\label{rmsx}
\end{figure}

Fig.~\ref{rmsx} (Dolag et al., in preparation) shows 
the $\sigma_{\rm RM}$-S$_{\rm X}$ plot for some more clusters, 
both regular and irregular.
The relation is still valid up to several orders of magnitude.
Therefore the reason why 
the magnetic fields at the center of relaxed system are higher 
than those in irregular clusters seems to be related
to a relation
between the magnetic field strength and the gas density.
The high magnetic field strength in the regular clusters are 
connected with the high
values of the gas density in the core of regular clusters. 

\subsection{Magnetic field power spectrum}
Contrary to what expected for a cluster magnetic field
model tangled on a single scale, many RM distributions
show clear evidence for
a non-zero $<$RM$>$ if averaged over areas comparable with the radio
source size.
This same property is observed for clusters located at
high galactic latitudes and in general cannot be entirely attributed
to the contribution from our own Milky Way Galaxy.  These  $<$RM$>$
offsets are likely due to fluctuations of the cluster magnetic fields on
scales greater than the sources size, i.e. considerably larger of those
responsible for the RM dispersion (Murgia et al. 2004).
The magnetic field must therefore be both tangled on sufficiently
small scales in order to produce the smallest structures observed in
the RM images and also fluctuate on scales one, or even two,
orders of magnitude larger.  For this reason, it is necessary to
consider cluster magnetic field models where both small and 
large-scales structures coexist.

Several efforts has been done to try to extend
the analysis of the RM maps beyond the tangled cell model.

En{\ss}lin \& Vogt (2003) and  Vogt \& En{\ss}lin (2003)
pointed out that the cluster magnetic field
model tangled on a single scale is not realistic because it does not
satisfy the condition $\rm\nabla$$^. \vec{B}=0$. 
They developed a Fourier analysis of RM images to 
determine the magnetic field power spectrum and therefore the magnetic field 
correlation length and strength.  
On the basis of their Fourier analysis, Vogt \& En{\ss}lin (2005)
applied a Bayesian view on the RM map of Hydra~A and
found a Kolmogoroff spectral index.
Their analysis, however, is limited to a narrow range of  
spatial scale and so it is insensitive to the
possible presence of magnetic fields on large scales.

\begin{figure}
\resizebox{\hsize}{!}
{\includegraphics[]{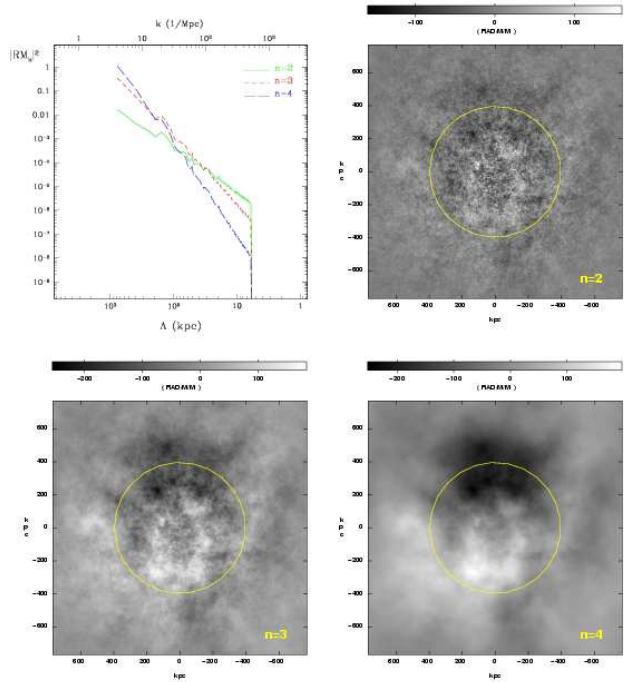}}
\caption{
Simulated cluster RM images for different values of the magnetic field power 
spectrum spectral index $n$ (Murgia et al. 2004). 
The three power spectra are normalized to have the 
same total magnetic field energy which is distributed over 
the range of spatial scales
 from 6 kpc up to 770 kpc. The average field at the cluster center 
is $<$B$>_{\rm0}$ $=1 ~\mu$G and its  
energy density decreases from the cluster center 
according to $B^{2}\propto n_{\rm e}(r)$, where 
$n_{\rm e}(r)$ is the gas density profile.
 Each RM image shows a field of view of about $1.5\times 1.5$ Mpc while the cluster core
 radius (indicated by the circle) is 400 kpc.
 The two-dimensional power spectra of the simulated RM images 
are shown in the top left panel.
 They have the same slope as their parent magnetic field power spectra and
 they span an equivalent range of spatial scales.
}
\label{faraday}
\end{figure}

Murgia et al. (2004)
simulated random 3D magnetic fields with an 
isotropic power-law
power spectrum\footnote{The power spectra 
are expressed as a vectorial form in $k$-space.
The one-dimensional forms can be obtained by multiplying by $4\pi k^2$
and $2\pi k$ respectively the three and the two-dimensional power spectra
} :
\begin{equation}
|B_{\kappa}|^2 \propto {\kappa}^{-n}
\label{power}
\end{equation}
\noindent
where ${\kappa}$ represents the wave number of the fluctuation scale.  They
investigated the effects of the Faraday rotation on the
polarization properties of radio galaxies and radio halos, by
analyzing the rotation measure produced
by a magnetic field with a
power spectrum which extends over a large range of spatial scales
and with different values of the spectral index $(n=2,3,4)$.
Fig.~\ref{faraday} shows the 
simulated RM images obtained with different values
of the index $n$ for a typical cluster of galaxies (see caption for
more details).
It is evident that the same cluster magnetic field energy
density will generate different magnetic field configurations with 
correspondingly different RM structures, for different values of the power
spectrum spectral index.
The resulting two-dimensional power spectra of the RM images are shown 
in the top-left panel of the figure.
They have the same slope as their parent's three-dimensional magnetic 
field power spectra and they cover an equivalent range of spatial scales. 

Murgia et al. (2004) compared the RM images,
obtained by Feretti et al. (1999), of three radio galaxies 
in the cluster A119 with the expectation
of a multi scale magnetic field model.
The modeled magnetic field is distributed 
over a wide range of spatial scales 
(6 - 770 kpc) and according to the $\sigma_{\rm RM}$-S$_{\rm X}$ 
relation (Dolag et al. 2001) it decreases from the cluster center
outward following the gas density ($B\propto n_{\rm e}^{0.9}$).
They found that the data are well describe with a
spectral index $n=2$ (flatter than a Kolmogoroff $n=11/3$)
and a central magnetic field strength $<$B$>_{\rm 0}~\simeq 5~\mu$G. 
The mean field strength within a spherical
volume of radius $r=3r_{\rm c}$  
gives $<$B$>~\simeq 1.5~\mu$G.

\section{Magnetic fields ordered on large scales}

While relics are usually strongly polarized 
no significant polarization has been
detected in general in radio halos (see e.g. A2256 by
Clarke \& En{\ss}lin this meeting).

The difficulty of detecting the polarized emission in halos 
has been interpreted as the result of two concurrent effects:
internal Faraday rotation and beam depolarization.
The thermal intracluster gas is mixed with the relativistic
plasma thus, due to internal Faraday rotation,
significant depolarization may occur within radio halos.
Moreover, as a consequence of their 
extremely low surface brightness, 
radio halos have been studied so 
far at low spatial resolution. This could result
in a significant decrease
of the observed fractional polarization if the cluster
magnetic field is tangled on scales smaller than the beam.

Murgia et al. (2004) pointed out that 
morphology and polarization information of radio halos
may provide important constrains on the power spectrum of 
the magnetic field fluctuations on large scales.
In particular their simulations shown that if the outer
scale of the magnetic field fluctuations extends up to
some hundred kpc, and if the power 
spectrum of the cluster magnetic field is relatively steep ($n\geq$3) 
there could be a chance
of detecting filamentary polarized emission in the radio halo if
observations are performed at sub-arcminute angular resolution.
They derived from RM images of cluster radio galaxies that
the magnetic field power spectrum of A2255 appears steep.
Therefore they observed A2255 with the purpose
of detecting polarized emission
from the radio halo and obtaining information on the degree of
ordering of the cluster magnetic field.
Fig.~\ref{halopol} displays the result of such observation.
The radio halo in A2255 show filamentary structures 
that are strongly polarized (Govoni et al. 2005).
The fractional linear polarization reaches levels of $\simeq$ 20$-$40\%
and the magnetic fields appear ordered on scales of $\sim$400 kpc.
This is the first successful attempt 
to image polarized emission from a radio halo and provides 
strong evidence that in this cluster the magnetic field 
is ordered on large scales.

A recent important development in the radio polarimetry is the
technique of RM-synthesis
(Brentjens \& de Bruyn 2005; de Bruyn \& Brentjens 2005,
see also this meeting).
Applied to the Perseus cluster it allowed the detection 
of Mpc-scale very faint polarized emission at the periphery
of the cluster.

Future observations with the new generation of radio telescopes 
(EVLA, LOFAR, LWA, SKA) will allow a detailed study of these 
faint extended structures to
improve our knowledge of the intensity and structure 
of large-scale magnetic fields.  

\begin{figure}
\resizebox{\hsize}{!}
{\includegraphics[]{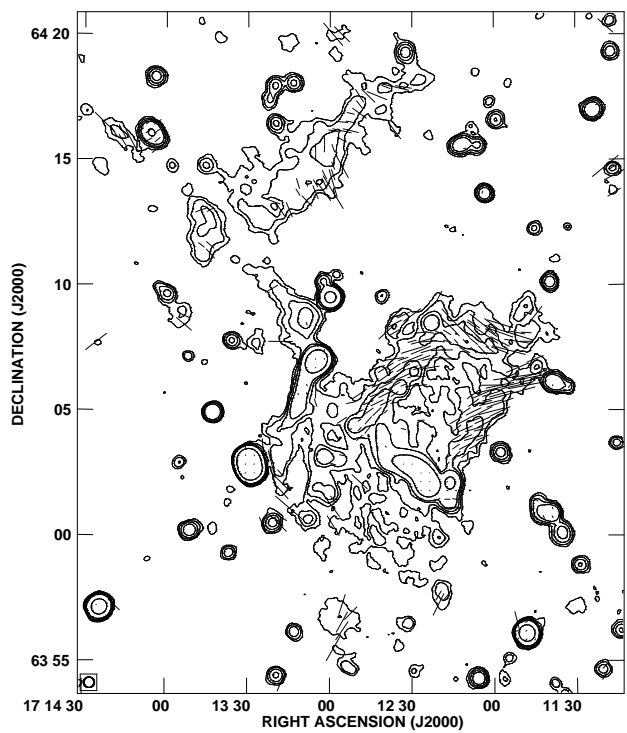}}
\label{halopol}
\caption{
Total intensity radio contours of A2255 at 1.4 GHz (Govoni et al. 2005). 
The contours of the total intensity 
are overlaid on the polarization vectors.
The vector orientation represents the projected E-field
while their length is proportional to the fractional polarization
(1\arcmin=50\%).
}
\label{halopol}
\end{figure}

\section{Conclusions}
Magnetic fields are common in regular and irregular clusters
and most of what we know comes from the detection of radio waves.

Our knowledge of the magnetic field properties in galaxy clusters 
has significantly improved in recent years, owing to the improved
capabilities of the instruments.

I have reviewed the main issues that have led our knowledge
on magnetic fields in clusters by focusing on studies of
cluster-wide diffuse sources
and from studies of the rotation measure of polarized radio
galaxies.
I have shown how it is possible to 
constrain the magnetic field strength and magnetic field
power spectrum, and I have discussed the relation between the 
magnetic field strength and the intracluster gas density.

From the results presented in the previous sections, it is derived 
that cluster magnetic field strengths obtained from RM arguments
are higher than the estimates
obtained from the diffuse synchrotron
radio halo emission under equipartition conditions.
However, both methods are based on many simplifying assumptions 
moreover RM gives the field along the line of
sight while the equipartition values refer to uniform 
magnetic field estimates averaged over large volumes.
The discrepancy may be alleviated by considering
that the magnetic field is not constant through the cluster, and shows
a complex structure. In particular, the magnetic field intensity
declines with the cluster radius with a dependence on the
thermal gas density.

One can device a scenario in which cluster magnetic field
fluctuate both on small scales (visible through RM of radio galaxies) 
and large scales (visible through cluster-wide diffuse sources). 
But only for those clusters for which
the power spectrum of the magnetic field fluctuations is steep 
enough polarized filamentary structures will be detectable.

%\begin{figure}
%\resizebox{\hsize}{!}
%{\includegraphics[]{figure.ps}}
%\caption{}
%\label{figlabel}
%\end{figure}

%\begin{table}[h]
%\caption{}
%\label{tablabel}
%\begin{tabular}{cc}\hline
%Quantity 1 & Quantity 2\\
%(unit1) & (unit2) \\
%\hline
%1 & 2 \\
%3 & 4\\
%\hline
%\end{tabular}
%\end{table}

\acknowledgements
I'm grateful to my collaborators Luigina Feretti, 
Gabriele Giovannini, Matteo Murgia, Emanuela Orr\'u
and Klaus Dolag for interesting discussions and helpful suggestions.
I wish to thank the organizers for inviting me to take part
in this great conference.

%\begin{appendix}
%\section{}
%\end{appendix}

\end{document}